\documentclass{emulateapj}

\shorttitle{Spin Equilibrium by APSWs}
\shortauthors{Matt \& Pudritz}

\begin{document}

\title{Accretion-Powered Stellar Winds III: Spin Equilibrium Solutions}


\author{Sean Matt\altaffilmark{1} and Ralph E. Pudritz\altaffilmark{2}}

\affil{$^1$Department of Astronomy, University of
Virginia, P.O. Box 400325, Charlottesville, VA 22904-4325; seanmatt@virginia.edu}

\affil{$^2$Physics and Astronomy Department, McMaster University,
Hamilton, ON L8S 4M1, Canada; pudritz@physics.mcmaster.ca}


\begin{abstract}


We compare the stellar wind torque calculated in a previous work
(Paper~II) to the spin-up and spin-down torques expected to arise from
the magnetic interaction between a slowly rotating ($\sim 10$\% of
breakup) pre-main-sequence star and its accretion disk.  This analysis
demonstrates that stellar winds can carry off orders of magnitude
more angular momentum than can be transferred to the disk, provided
that the mass outflow rates are greater than the solar wind.  Thus,
the equilibrium spin state is simply characterized by a balance
between the angular momentum deposited by accretion and that extracted
by a stellar wind.  We derive a semi-analytic formula for predicting
the equilibrium spin rate as a function only of the ratio of $\dot
M_{\rm w} / \dot M_{\rm a}$ and a dimensionless magnetization
parameter, $\Psi \equiv B_*^2 R_*^2 (\dot M_{\rm a} v_{\rm
  esc})^{-1}$, where $\dot M_{\rm w}$ is the stellar wind mass outflow
rate, $\dot M_{\rm a}$ the accretion rate, $B_*$ the stellar surface
magnetic field strength, $R_*$ the stellar radius, and $v_{\rm esc}$
the surface escape speed.  For parameters typical of accreting
pre-main-sequence stars, this explains spin rates of $\sim 10$\% of
breakup speed for $\dot M_{\rm w} / \dot M_{\rm a} \sim 0.1$.
Finally, the assumption that the stellar wind is driven by a fraction
of the accretion power leads to an upper limit to the mass flow ratio
of $\dot M_{\rm w} / \dot M_{\rm a} \la 0.6$.

\end{abstract}


\section{Introduction} \label{sec_intro}

The slow rotation rates of low to intermediate mass ($\la 2 M_\odot$)
pre-main-sequence stars remains one of the most important aspects of
star formation that has, so far, resisted a generally accepted
explanation.  By the time they become optically visible as T Tauri
stars \citep[TTSs;][]{joy45}, approximately half of them are observed
to rotate at approximately 10\% of breakup speed \citep[the ``slow
rotators''; e.g.,][]{vogelkuhi81, bouvier3ea97, rebull3ea04,
herbstea07}.  This is a surprise because many TTSs (the Classical T
Tauri stars; CTTSs) are actively accreting material from surrounding
Keplerian disks \citep{lyndenbellpringle74, bertout3ea88,
calvetgullbring98, muzerolle3ea01}.  At a typical accretion rate of
$\dot M_{\rm a} \sim 10^{-8} M_\odot$ yr$^{-1}$, the angular momentum
deposited by accreting disk material should spin up a CTTS to near
breakup speed in $\sim 10^6$ years \citep{hartmannstauffer89,
mattpudritz07coolstars}.  Since the accretion phase lasts for $10^6$
-- $10^7$ years \citep{lyolawson05, jayawardhanaea06}, since the stars
accrete at much higher rates prior to the TTS phase, and since the
stars are still contracting \citep{rebullea02}, an efficient angular
momentum loss mechanism is required to explain the existence of the
slow rotators.

A few interesting and important ideas for explaining the TTS slow
rotators have been developed over the last two decades.  These have
resulted in the star-disk interaction model of \citet{ghoshlamb78},
applied to CTTSs by \citet[][and see \citealp{camenzind90}]{konigl91},
the X-wind model \citep{shuea94}, and the idea that stellar winds
provide strong torques \citep{hartmannstauffer89, toutpringle92,
paatzcamenzind96, ferreira3ea00, mattpudritz05l}.  Although both have
advanced our understanding of the magnetic star-disk interaction,
neither the Ghosh \& Lamb nor X-wind models are without problems
\citep{ferreira3ea00, uzdensky04, mattpudritz05}, and the idea that
stellar winds are important has not yet been worked out in sufficient
detail to compare to the other models.


In \citet[][hereafter Paper~I]{mattpudritz05l}, we further explored
powerful stellar winds as a solution to the angular momentum problem
and suggested that a fraction of the accretion power provides the
energy necessary to drive the wind.  We showed that stellar winds are
capable of carrying off the accreted angular momentum, provided that
$\dot M_{\rm w} / \dot M_{\rm a} \sim 0.1$, where $\dot M_{\rm w}$ is
the outflow rate of material that is magnetically connected to the
star (the ``stellar wind'').  This analysis included a formulation for
the stellar wind torque that contained the Alfv\'en radius ($r_{\rm
  A}$), which is not easily determined a priori in the wind, and the
conclusions were based on a one-dimensional scaling estimate of this
important physical quantity.  Thus, while it is clear that
accretion-powered stellar winds (APSWs) can in principle provide the
necessary spin-down torque, this idea requires further development to
produce a more detailed model.

Toward this goal, \citet[][hereafter Paper~II]{mattpudritz08II} used
2-dimensional (axisymmetric) magnetohydrodynamic simulations to solve
for $r_{\rm A}$ and calculate realistic stellar wind torques for a
range of parameters.  In the present paper, we use the stellar wind
solutions of Paper~II to compare the stellar wind torque to the
torques expected to arise from the star-disk interaction.
Furthermore, we find new solutions for stellar spins, based upon
torque balance between the accretion torque and the APSW spin-down
torque.  This paper begins with a brief description of the simulation
results of Paper~II (\S \ref{sec_simulations}).  We then compare the
stellar wind torque to the star-disk spin-down torque in section
\ref{sec_tdsd} and then to the star-disk spin-up torque in section
\ref{sec_equilibrium}, which contains spin-equilibrium solutions.
Section \ref{sec_discussion} contains a summary and discussion.

\section{Results of Stellar Wind Simulations} \label{sec_simulations}

This section contains a brief description of the simulation results of
Paper~II that we will use for our analysis, and the reader will find
details in that paper.  The primary purpose of the simulations was to
compute the spin-down torque on a star, due to the angular momentum
outflow in a wind.  We used numerical magnetohydrodynamic simulations
to directly calculate the torque $\tau_{\rm w}$ from steady-state, 2D
(axisymmetric) winds from isolated stars.  We adopted coronal
(thermal-pressure-driven) winds as a proxy for the unknown wind
driving mechanism.  In the simulations, the torque is entirely
determined by the seven key parameters listed in table
\ref{tab_parms}.  These are the stellar mass, $M_*$; stellar radius,
$R_*$; strength of the rotation-axis-aligned dipole magnetic field at
the surface and equator of the star, $B_*$; spin rate expressed as a
fraction of breakup speed,
\begin{eqnarray}
\label{eqn_f}
f \equiv \Omega_* R_*^{3/2} (G M_*)^{-1/2},
\end{eqnarray}
where $\Omega_*$ is the angular spin rate of the star; mass outflow
rate in the stellar wind, $\dot M_{\rm w}$; ratio of the thermal sound
speed to the escape speed, evaluated at the base of the wind (just
above the stellar surface), $c_{\rm s} / v_{\rm esc}$; and adiabatic
index, $\gamma$.

\begin{deluxetable}{ll}
\tablewidth{0pt}
\tablecaption{Fiducial Stellar Wind Parameters \label{tab_parms}}
\tablehead{
\colhead{Parameter} &
\colhead{Value}
}

\startdata

$M_*$                     & 0.5 $M_\odot$ \\
$R_*$                     & 2.0 $R_\odot$ \\
$B_*$ (dipole)            & 200 G         \\
$f$                       & 0.1           \\
$\dot M_{\rm w}$          & $1.9 \times 10^{-9} M_\odot$ yr$^{-1}$ \\
$c_{\rm s} / v_{\rm esc}$ & 0.222          \\
$\gamma$                  & 1.05           \\

\enddata


\end{deluxetable}

Table \ref{tab_parms} lists the value of each parameter adopted for a
fiducial case.  Paper~II contained a parameter study in which each of
the seven parameters were varied relative to the fiducial case, and 14
cases from the parameter study are listed in table
\ref{tab_torques}\footnote{In this paper, we do not discuss the cases
from Paper~II that include a quadrupole magnetic field, nor the
extremely slow rotator case (with $f = 0.004$).}.  In each case, six
of the parameters were held fixed at the fiducial value (as given in
table \ref{tab_parms}), and one parameter was varied as indicated by
the first column of table \ref{tab_torques}.

To compare with analytic theory, we also calculated the effective Alfv\'en
radius ($r_{\rm A}$), where the poloidal wind velocity equals the
poloidal Alfv\'en speed, using an analytic formula for the stellar
wind torque,
\begin{eqnarray}
\label{eqn_tw}
\tau_{\rm w} = - \dot M_{\rm w} \Omega_* \left< r_{\rm A}^2 \right>.
\end{eqnarray}
Since our simulations are multi-dimensional, we have used $\left<
r_{\rm A}^2 \right>$, which is the mass-loss-weighted average of
$r_{\rm A}^2$.  Hereafter, we'll refer to $\left< r_{\rm A}^2
\right>^{1/2}$ generically as $r_{\rm A}$.  Using the simulation
result for $\tau_{\rm w}$, equation \ref{eqn_tw} defines the value of
$r_{\rm A}$, which is tabulated for all cases in the second column of
table \ref{tab_torques}.

In this paper, we make use of the semi-analytic
formulation for the Alfv\'en radius from Paper~II,
\begin{eqnarray}
\label{eqn_rasim}
{r_{\rm A} \over R_*} = K
  \left({{B_*^2 R_*^2} \over {\dot M_{\rm w} v_{\rm esc}}}\right)^m,
\end{eqnarray}
where $K$ and $m$ are dimensionless constants fit to the simulation,
and $v_{\rm esc} = (2 G M_* / R_*)^{1/2}$ is the escape speed from the
stellar surface.  Paper~II showed that the values of $K \approx 2.11$
and $m \approx 0.223$ well-describe (to better than 1\%) the fiducial
case and those eight other cases with variations on $B_*$, $R_*$,
$\dot M_{\rm w}$, and $M_*$.  Although this is only approximately
valid for situations with different wind acceleration rates or
different rotation rates (in which case the values of $K$ and $m$ are
slightly different; see Paper~II), the formulation of equation
\ref{eqn_rasim} serves well as an indication of the approximate
dependence of the stellar wind on parameters, which will be important
for discussing a wide range of possible conditions.

The form of equation (\ref{eqn_rasim}) is similar to that derived by
(e.g.)\ \citet[][and see \citealp{ferreira97}]{pelletierpudritz92} for
the general theory of centrifugally driven disk winds.  The quantity
in brackets measures the magnetization of the wind.  By assuming that
the Alfv\'en speed $v_{r,A}$ (at the Alfv\'en radius) is directly
proportional to $\Omega_*r_A$, a relation of the kind given by
equation (\ref{eqn_rasim}) can be derived \citep[e.g., see equation
  2.27 of][]{pelletierpudritz92}.  In that case, the value of the
index is $m=1/3$.  While this value is not far from the results of our
numerical simulations, the difference is significant.  One key reason
for this may be that disk winds are in the regime of so-called fast
magnetic rotators, whereas the rather slowly rotating TTS are either
slow magnetic rotators (where wind-driving forces dominate over
centrifugal ones) or are intermediate between these two regimes (see
Paper~II).

For the discussion that follows, it is useful to highlight how the
lever arm ($r_{\rm A}$) and wind torque responds to changing the mass
load (the mass loss rate) of the wind.  The fact that the Alfv\'en
lever arm in a hydromagnetic wind gets smaller as the mass load of the
outflow increases, as is seen in equation (\ref{eqn_rasim}), seems to
suggest that the wind would become ineffective.  This is certainly not
true however, because equation (\ref{eqn_tw}) assures that an increase
in wind mass loss rate leads to a net increase in the torque that the
wind exerts upon the star (the net wind torque scales as $\dot
M_w^{1-2m}$).  This is the basic reason why, by having an outflow rate
that is a substantial fraction of the accretion rate, an
accretion-powered stellar wind can be effective in countering the
accretion torque.


\begin{deluxetable}{lcccccc}
\tablewidth{0pt}
\tablecaption{Stellar Wind Alfv\'en Radii and Comparison to Star-Disk Spin-Down Torques \label{tab_torques}}
\tablehead{
\colhead{Case} &
\colhead{$r_{\rm A} / R_*$} &
\colhead{$\tau_{\rm w} / \tau_{\rm dsd}$} &
\colhead{$\tau_{\rm w} / \tau_{\rm dsd}$} \\
\colhead{} &
\colhead{} &
\colhead{$(\beta = 0.1)$} &
\colhead{$(\beta = 0.01)$} 
}



\startdata

fiducial                         &  6.97  &  59  &  490  \\
$f$ = 0.2                        &  6.26  &  23  &  200  \\
$f$ = 0.05                       &  7.65  &  140 &  1200 \\
$B_*$ = 400 G                    &  9.55  &  27  &  230  \\
$B_*$ = 2 kG                     &  19.3  &  4.6 &  39   \\
low $\dot M_{\rm w}$$^a$         &  11.8  &  17  &  140  \\
very low $\dot M_{\rm w}$$^a$    &  16.7  &  6.7 &  57  \\
$R_*$ = 1.5 $R_\odot$            &  5.96  &  86  &  730  \\
$R_*$ = 3 $R_\odot$              &  8.75  &  34  &  280  \\
$M_*$ = 0.25 $M_\odot$           &  7.52  &  49  &  410  \\
$M_*$ = 1 $M_\odot$              &  6.42  &  70  &  590  \\
$c_{\rm s}/v_{\rm esc}$ = 0.245  &  6.64  &  53  &  440  \\
$c_{\rm s}/v_{\rm esc}$ = 0.192  &  7.23  &  63  &  530  \\
$\gamma$ = 1.10                  &  7.79  &  72  &  610  \\

\enddata

\footnotetext{$^a$ The mass outflow rate in the low and very low $\dot
M_{\rm w}$ cases is $1.9 \times 10^{-10}$ and $3.8 \times 10^{-11}
M_\odot$ yr$^{-1}$, respectively.}

\end{deluxetable}

It is our goal here to compare the stellar wind torque to the torque
expected to arise from the star-disk interaction, and the latter has
only been determined thus far for a dipolar geometry.  So we only
consider here the cases from Paper~II with a dipole magnetic field.
We also adopt the following assumptions.  Paper~II indicated that the
details of the wind driving have a relatively small, but not entirely
negligible, effect on the stellar wind torque.  In the absence of a
detailed model for how APSWs are driven, we assume that the velocity
profile of an APSW does not differ substantially from our simulations
(Paper~II), so that the calculated torques are valid.  Secondly,
Paper~II considered winds from isolated stars.  Here, we will use the
computed torques to develop the APSW scenario, in which stellar winds
are accompanied by disk winds, accretion flows, and the general
star-disk interaction (see figure 1 of Paper~I).  In reality, the
accretion disk blocks a portion of the stellar wind, and it is not
clear how much this will affect the stellar wind torque.  For the
present study, we will assume that the presence of the disk and
accretion will not significantly influence the stellar wind torques as
computed in Paper~II.




\section{Stellar Wind vs.\ Star-Disk Spin-Down Torque} \label{sec_tdsd}

The magnetic interaction between the star and disk results in angular
momentum transfer between the two.  All models that calculate the
torque on the star from this interaction are based on the framework
constructed by \citet{ghoshlamb78}.  In this general model, some of the
stellar magnetic dipole flux connects to the accretion disk and
conveys torques between the star and disk.  The net torque can be
separated into a spin-up part that adds angular momentum to the star
and a spin-down part that removes angular momentum from the star,
giving it back to the disk.  In the absence of a stellar wind, a
spin-down torque only arises when there is a magnetic connection
between the star and the region of the disk outside the corotation
radius,
\begin{eqnarray}
\label{eqn_rco}
R_{\rm co} \equiv f^{-2/3} R_*.
\end{eqnarray}
It is assumed that the disk is capable of transporting away the excess
angular momentum it receives from the star.  The goal of this section
is to compare the stellar wind torque to the spin-down torque arising
from the star-disk magnetic connection, to determine under which
circumstances each of these torques may be important and aid in the
angular momentum loss from the star.

To calculate the star-disk spin-down torque, $\tau_{\rm dsd}$, we
follow \citet[][hereafter MP05b]{mattpudritz05}, who formulated a
Ghosh \& Lamb type model that includes the effect of the opening of
magnetic field lines via the differential rotation between the star
and disk.  In this case, $\tau_{\rm dsd}$ is calculated by considering
only the magnetic flux that remains closed (connected), parametrized
as having an azimuthal twist of less than a critical angle,
$\tan^{-1}(\gamma_{\rm c})$.  Here we adopt the value of $\gamma_{\rm
c} = 1$ suggested by \citet{uzdensky3ea02}.  By combining equation (9)
and (22) of MP05b with equation (\ref{eqn_rco}), we find
\begin{eqnarray}
\label{eqn_tdsd}
\tau_{\rm dsd} = - {\chi(\beta) \over 3} f^2 B_*^2 R_*^3,
%
\end{eqnarray}
where 
\begin{eqnarray}
\label{eqn_chi}
\chi(\beta) \equiv \beta^{-1} [1 + (1+\beta)^{-2} - 2(1+\beta)^{-1}]
\end{eqnarray}
is a dimensionless function of the strength of the effective magnetic
diffusion rate in the disk.  This is parametrized by $\beta$, which
for a standard $\alpha$-disk \citep{shakurasunyaev73} is
$\beta$~$\equiv$~$\alpha h (P_{\rm t} r)^{-1}$, where $\alpha$ has its
usual meaning, $h$ is the disk scale height at radius $r$, and $P_{\rm
t}$ is the turbulent magnetic Prandtl number\footnote{Note that this
$\beta$ has no relation to the usual ``plasma beta'' parameter that
often appears in MHD studies.}.  Small values of $\beta$ correspond to
strong coupling, and MP05b suggested that $\beta \sim 0.01$ was
appropriate for real disks.  For strong coupling (small $\beta$), the
magnetic field will be highly twisted azimuthally, leading to more
open flux and a weaker star-disk spin-down torque.  For small $\beta$,
$\chi(\beta) \approx \beta$, but $\chi(\beta)$ has a maximum value of
0.25 when $\beta = 1$ (e.g., see fig.\ 7 of MP05b).

Equation (\ref{eqn_tdsd}) indicates that the star-disk spin-down
torque is completely independent of the accretion rate.  This torque
only requires that there exists a Keplerian disk outside $R_{\rm co}$,
to which the star can connect, and it does not matter whether or not
there is net accretion onto the star.  The dependence on the stellar
spin rate $f$ is due to the fact that when the star spins faster,
$R_{\rm co}$ is closer to the star, where the magnetic field is
stronger.

We compare the stellar wind torque computed in our simulations to the
star-disk spin-down torque by listing the ratio $\tau_{\rm w} /
\tau_{\rm dsd}$ in the last two columns of table \ref{tab_torques}.
We consider both a case with $\beta = 0.1$, resulting in $\chi \approx
0.0826$, and a case with $\beta = 0.01$, resulting in $\chi \approx
0.00980$.  It is apparent that the simulated stellar wind torques are
tens to hundreds of times greater than $\tau_{\rm dsd}$, the larger
difference existing for smaller values of $\beta$.

Thus, for the simulated winds, we see empirically that the stellar
wind is much more effective at spinning down the star than is the
star-disk connection.  This can be understood qualitatively as
follows.  In the stellar magnetosphere, any torque on the star is
primarily conveyed by the azimuthal twisting of its magnetic field.
In the case of the magnetic field connecting the star to the disk,
there is a limit to how much the magnetic field can be twisted before
the connection is lost \citep[][MP05b]{uzdensky3ea02}.  In the case of
a stellar wind flowing along the field, there is no such limit on the
twist.  The larger the mass outflow rate in the wind, the less capable
is the magnetic field to keep wind material corotating with the star,
and so the larger will be the twist of the field.

To show the dominance of the stellar wind torque more generally and to
identify the circumstances under which it might not dominate, we will
use the semi-analytic formulation of equation (\ref{eqn_rasim}).  By
combining equations (\ref{eqn_f}) -- (\ref{eqn_rasim}), one obtains
\begin{eqnarray}
\label{eqn_twind2}
\tau_{\rm w} = - {K^{1/m} \over \sqrt{2}}
     \left({R_* \over r_{\rm A}}\right)^{1/m - 2} f B_*^2 R_*^3
\end{eqnarray}
for the stellar wind torque.  At first it may seem unusual that
$\tau_{\rm w}$ is weaker when the magnetic lever arm length, $r_{\rm
  A}$, is larger (for fixed $f B_*^2 R_*^3$).  However this simply
indicates that the stellar wind torque increases with increasing $\dot
M_{\rm w}$, as noted in section \ref{sec_simulations}.  Also, note
that a weak $f$-dependence of $r_{\rm A}$ is not characterized in
equation (\ref{eqn_rasim}) (see Paper~II), so the exact dependence of
the torque on $f$ is not captured in equation (\ref{eqn_twind2}).

\begin{figure}
\epsscale{1.15}
\plotone{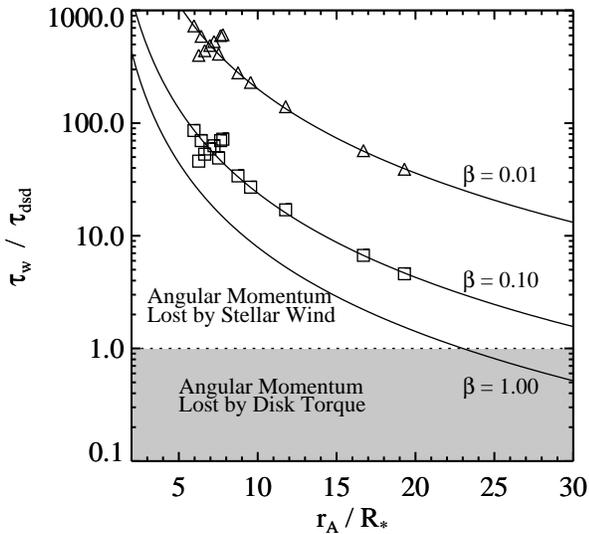}

\caption{The ratio of the stellar wind torque to the spin-down portion
of the star-disk interaction torque versus the magnetic lever arm
length in the stellar wind.  When the ratio is much greater than one,
the stellar wind is most important for angular momentum loss from the
star.  The lines correspond to equation (\ref{eqn_ratio}), assuming a
stellar spin rate of $f = 0.1$ and three different values of the
factor $\chi(\beta)$ (see text), corresponding to $\beta$ = 0.01, 0.1,
and 1, as indicated.  The values listed in table \ref{tab_torques},
obtained by comparing simulated wind torques to the analytic disk
torques, are plotted as squares (for $\beta = 0.1$) and triangles
($\beta = 0.01$).  The figure indicates that, unless the magnetic
lever arm length is very long (e.g., for very low stellar wind mass
loss rate), the spin-down torque from the disk is negligible.
\label{fig_torqueratio}}

\end{figure}

By combining equations (\ref{eqn_tdsd}) and (\ref{eqn_twind2}), using
$K = 2.11$ and $m = 0.223$ (see \S \ref{sec_simulations}), one finds
\begin{eqnarray}
\label{eqn_ratio}
{\tau_{\rm w} \over \tau_{\rm dsd}} \approx 6.0 \times 10^{4}
      \left({\chi(\beta) \over 10^{-2}}\right)^{-1} 
      \left({f \over 0.1}\right)^{-1} 
      \left({R_* \over r_{\rm A}}\right)^{2.48}.
\end{eqnarray}
The lines in figure \ref{fig_torqueratio} show equation
(\ref{eqn_ratio}) for $f=0.1$ and for three different values of
$\beta$.  This includes a line for $\beta=1$, which corresponds to the
strongest possible star-disk spin-down torque (as discussed by MP05b).
Even in this case, the fiducial stellar wind torque is $\sim 20$ times
stronger than the star-disk spin-down torque.  For smaller, more
realistic values of $\beta$, the star-disk spin-down torque only
becomes weaker, while the stellar wind torque is not affected.  It is
clear that for the parameters considered here, the angular momentum
extracted by the stellar wind completely dominates over that which can
be transferred from the star to the disk.

The stellar wind torque becomes weaker relative to $\tau_{\rm dsd}$
when $r_{\rm A}$ is larger (e.g., for smaller $\dot M_{\rm w}$) or for
more rapidly spinning stars (larger $f$).  For the case of $\beta =
0.01$, favored by MP05b, $\tau_{\rm w}$ will be larger than $\tau_{\rm
  dsd}$ for a star with $f=0.1$, as long as $r_{\rm A} \la 84 R_*$.
This limiting value is much longer than any of the lever arm lengths
listed in table \ref{tab_torques}.  As an example, for all else being
equal to the fiducial case, equation (\ref{eqn_rasim}) suggests that
$\tau_{\rm w} > \tau_{\rm dsd}$, as long as $\dot M_{\rm w} \ga 3
\times 10^{-14} M_\odot$~yr$^{-1}$.  This limit is comparable to the
solar wind mass loss rate.  If the stellar dipole field strength is
instead $B_*=2$ kG, the limit becomes $\dot M_{\rm w} \ga 3 \times
10^{-12} M_\odot$.

The squares and triangles in figure \ref{fig_torqueratio} represent
the data from table \ref{tab_torques}.  Note that nine of the data
points (for each $\beta$) match the line very well.  This is expected
since the results from these cases were used to obtain the value of
$K$ and $m$ used in equation (\ref{eqn_ratio}).  There are five cases
(for each $\beta$) that lie slightly off of the line.  Three of them
represent the last three cases listed in table \ref{tab_torques}, and
the other two are the $f=0.2$ and $f=0.05$ cases (which, for the plot,
we have scaled by a factor of $f/0.1$, to take into account the spin
dependence of equation (\ref{eqn_ratio})).  These five cases are not
expected to match exactly since equation (\ref{eqn_rasim}) is not
precise for cases with different wind driving or spin rates than the
fiducial case (see Paper~II).  Thus, the scatter of these five cases
around the line indicates a sort of uncertainty of the semi-analytic
formula for the stellar wind, due to variations in the wind driving
mechanism and stellar spin rate.  It is evident from the figure that
this uncertainty does not affect the main conclusion here that stellar
wind torques dominate the spin down of the star.

Thus, for the slow rotators ($f \sim 0.1$), we conclude that a stellar
wind will transport much more angular momentum from the star than will
a magnetic connection to the disk, as long as the stellar wind mass
outflow rate is substantially larger than the solar wind mass outflow
rate.  For the systems considered here, with $\do M_{\rm w} \sim
10^{-9} M_\odot$~yr$^{-1}$, the stellar wind torque completely
dominates over any other spin-down torque felt by the star.  Since
$\tau_{\rm dsd}$ is negligible, the only important torques on these
stars are the spin-down torque from the stellar wind and the spin-up
portion of the star-disk interaction torque.  We compare these two
torques in the following section.


\section{Spin Equilibrium by an APSW} \label{sec_equilibrium}

Section \ref{sec_tdsd} revealed that, for the slow rotators with
substantial stellar winds, the spin-down torque felt by the star from
the star-disk interaction is negligible.  Thus, the spin state of the
star is characterized as a competition between the spin-up component
of the star-disk interaction torque and the spin-down by the stellar
wind.  If a system's parameters are measured, the theory can be used
to determine the net torque on the star.

Given enough time ($\sim 10^5$--$10^6$ yr for CTTSs), the stellar spin
should approach an equilibrium spin state in which the net torque on
the star is zero \citep[see, e.g.,][]{cameroncampbell93,
  armitageclarke96}.  The variability observed in accreting systems
\citep[e.g.,][]{hartmann97} suggests that spin equilibrium may only
represent a time-averaged state, and the condition of net zero torque
simply identifies where the net torque changes sign.  In any case, it
is instructive to examine the conditions of spin-equilibrium.

In sections \ref{sec_balancing} -- \ref{sec_power}, we examine the
expected spin-equilibrium state of the specific cases of stellar winds
simulated in Paper~II.  In section \ref{sec_semianalytic}, we use the
semi-analytic formulation of equation \ref{eqn_rasim} to make more
general conclusions.

     \subsection{Spin Equilibrium for Specific Cases} \label{sec_balancing}

The spin-up portion of the star-disk interaction torque comes
primarily from the accretion of material from the innermost part of
the disk onto the star.  In this section, we examine some specific
cases of spin equilibrium, by determining under what conditions this
spin-up torque balances the stellar wind torques for the simulations
listed in table \ref{tab_torques}.

When a star's magnetic field is strong enough, it will disrupt the
Keplerian disk at some radius, $R_{\rm t}$, the disk truncation
radius.  We calculate the location of $R_{\rm t}$ using the method and
equations contained in the Appendix, which follows MP05b.  From
$R_{\rm t}$, accreting material is channeled by the magnetic field to
the surface of the star.  There is a torque associated with the
truncation of the disk and the accretion of material from $R_{\rm t}$.
This torque, hereafter the ``accretion torque,'' is given by MP05b as
\begin{eqnarray}
\label{eqn_ta}
\tau_{\rm a} = \dot M_{\rm a} \sqrt{G M_* R_*} 
     \left[{\left({R_{\rm t} \over R_*}\right)^{1/2} - k^2 f}\right],
\end{eqnarray}
where $\dot M_{\rm a}$ is the mass accretion rate onto the stellar
surface and $k$ is the normalized radius of gyration of the star (we
assume $k^2 \approx 0.2$; \citealp{armitageclarke96}).  Equation
(\ref{eqn_ta}) assumes that all of the Keplerian specific angular
momentum of the disk material near $R_{\rm t}$ is transferred to the
star.  This naturally follows from the dynamical truncation of the
disk (e.g., \citealp{yi95, wang95}; MP05b) and is supported by
numerical simulations \citep[e.g.,][]{romanovaea02, long3ea05}.  It is
clear that the accretion torque depends both on $\dot M_{\rm a}$ and
$R_{\rm t}$.  At the same time, as shown in the Appendix, the location
of $R_{\rm t}$ itself depends on most of the parameters, including
$\dot M_{\rm a}$ and $\beta$.


\begin{deluxetable}{lcccc} 
\tablewidth{0pt}
\tablecaption{Spin-Equilibrium Results for $\gamma_{\rm c} = 1$ \& 
              $\beta = 0.1$ \label{tab_equilib0.1}}
\tablehead{
\colhead{Case} &
\colhead{$\dot M_{\rm w} / \dot M_{\rm a}$} &
\colhead{$R_{\rm t} / R_*$} &
\colhead{State} &
\colhead{$\epsilon_\infty$}
}



\startdata

fiducial                         &  0.43  &  4.4  &  2  &  0.71   \\
$f$ = 0.2                        &  0.21  &  2.9  &  2  &  0.39   \\
$f$ = 0.05                       &  0.83  &  5.9  &  1  &  1.2    \\
$B_*$ = 400 G                    &  0.23  &  4.5  &  2  &  0.37   \\
$B_*$ = 2 kG                     &  0.057 &  4.6  &  2  &  0.055  \\
low $\dot M_{\rm w}$             &  0.15  &  4.6  &  2  &  0.23   \\
very low $\dot M_{\rm w}$        &  0.076 &  4.6  &  2  &  0.094  \\
$R_*$ = 1.5 $R_\odot$            &  0.58  &  4.3  &  2  &  0.96   \\
$R_*$ = 3 $R_\odot$              &  0.28  &  4.5  &  2  &  0.46   \\
$M_*$ = 0.25 $M_\odot$           &  0.37  &  4.5  &  2  &  0.61   \\
$M_*$ = 1 $M_\odot$              &  0.51  &  4.4  &  2  &  0.84   \\
$c_{\rm s}/v_{\rm esc}$ = 0.245  &  0.48  &  4.4  &  2  &  0.87   \\
$c_{\rm s}/v_{\rm esc}$ = 0.192  &  0.40  &  4.4  &  2  &  0.61   \\
$\gamma$ = 1.10                  &  0.34  &  4.4  &  2  &  0.48   \\

\enddata

\end{deluxetable}

The spin equilibrium state is defined by 
\begin{eqnarray}
\label{eqn_teq}
\tau_{\rm a} = - \tau_{\rm w}.
\end{eqnarray}
Each of our wind simulation cases represents a specific set of values
for $\tau_{\rm w}$, $\dot M_{\rm w}$, $B_*$, $M_*$, $R_*$, and
$\Omega_*$.  For each simulation case, we used equation (\ref{eqn_ta})
and the method in the Appendix to determine the values of $R_{\rm t}$
and $\dot M_{\rm a}$ such that the condition (\ref{eqn_teq}) is
satisfied. We consider both a case with $\beta = 0.1$ and a case with
$\beta = 0.01$ (see \S \ref{sec_tdsd}).  The results, given as $\dot
M_{\rm w} / \dot M_{\rm a}$ and $R_{\rm t} / R_*$, are listed in the
2nd and 3rd columns of tables \ref{tab_equilib0.1} (for $\beta = 0.1$)
and \ref{tab_equilib0.01} (for $\beta = 0.01$).

A comparison between tables \ref{tab_equilib0.1} and
\ref{tab_equilib0.01} reveals that the disk magnetic coupling
parameter $\beta$ has little influence on the equilibrium values of
$\dot M_{\rm a}$ and $R_{\rm t}$.  This demonstrates that, although
$\beta$ has a large influence on the (negligible) spin-down part of
the star-disk interaction torque (as shown in \S \ref{sec_tdsd}),
$\beta$ has very little influence on the spin-up part.

For the specific cases of the simulated stellar winds, it is clear
that the equilibrium spin state is characterized by $\dot M_{\rm w} /
\dot M_{\rm a}$ of typically a few tens of percent.  This ratio is
smaller for cases with larger $r_{\rm A}$ (e.g., for larger field
strength or smaller $\dot M_{\rm w}$).  Thus, the cases listed in the
tables confirm the general conclusion of Paper~I, and represent valid
torque solutions for the spin equilibrium state.

     \subsection{Magnetic Connection State of the System} \label{sec_state}

As pointed out by MP05b (and see the Appendix), an accreting system
may exist in a state where the stellar magnetic field connects to the
disk outside the corotation radius, which they call ``state 2,'' and
which we implicitly assumed in section \ref{sec_tdsd}.  On the other
hand, if the disk truncation radius is sufficiently smaller than
$R_{\rm co}$, the star can lose its magnetic connection to all but the
very inner edge of the disk, which they call ``state 1.''  The
determination of $R_{\rm t}$ is different in the two states.  In the
absence of a stellar wind torque, a star in spin equilibrium must be
characterized by state 2 (MP05b).  Thus, having $R_{\rm t}$ very close
to $R_{\rm co}$ is a requirement of the ``disk locking'' models
\citep{konigl91, ostrikershu95, wang95}.  By contrast, this is not a
requirement of the APSW scenario.

\begin{deluxetable}{lcccc} 
\tablewidth{0pt}
\tablecaption{Spin-Equilibrium Results for $\gamma_{\rm c} = 1$ \& 
              $\beta = 0.01$ \label{tab_equilib0.01}}
\tablehead{
\colhead{Case} &
\colhead{$\dot M_{\rm w} / \dot M_{\rm a}$} &
\colhead{$R_{\rm t} / R_*$} &
\colhead{State} &
\colhead{$\epsilon_\infty$} 
}



\startdata

fiducial                         &  0.44  &  4.6  &  2  &  0.73   \\
$f$ = 0.2                        &  0.21  &  2.9  &  2  &  0.39   \\
$f$ = 0.05                       &  0.83  &  5.9  &  1  &  1.2    \\
$B_*$ = 400 G                    &  0.24  &  4.6  &  2  &  0.39   \\
$B_*$ = 2 kG                     &  0.057 &  4.6  &  2  &  0.055  \\
low $\dot M_{\rm w}$             &  0.15  &  4.6  &  2  &  0.23   \\
very low $\dot M_{\rm w}$        &  0.077 &  4.6  &  2  &  0.096  \\
$R_*$ = 1.5 $R_\odot$            &  0.59  &  4.4  &  1  &  0.98   \\
$R_*$ = 3 $R_\odot$              &  0.28  &  4.6  &  2  &  0.46   \\
$M_*$ = 0.25 $M_\odot$           &  0.38  &  4.6  &  2  &  0.63   \\
$M_*$ = 1 $M_\odot$              &  0.52  &  4.6  &  2  &  0.86   \\
$c_{\rm s}/v_{\rm esc}$ = 0.245  &  0.49  &  4.6  &  2  &  0.88   \\
$c_{\rm s}/v_{\rm esc}$ = 0.192  &  0.41  &  4.6  &  2  &  0.62   \\
$\gamma$ = 1.10                  &  0.35  &  4.6  &  2  &  0.49   \\

\enddata

\end{deluxetable}

Following MP05b (using equation \ref{eqn_trans}), we determined the
magnetic connection state of the spin-equilibrium systems described
above and listed this in the 4th column of tables \ref{tab_equilib0.1}
and \ref{tab_equilib0.01}.  Note that MP05b only consider a loss of
magnetic connection via the differential twisting of field lines.  The
stellar wind should also influence the connectedness between the star
and disk \citep{safier98}, but we do not attempt to quantify this
here.

Tables \ref{tab_equilib0.1} and \ref{tab_equilib0.01} reveal that most
(though not all) of the simulated cases are in a magnetic connection
state 2, while in spin-equilibrium.  A characteristic of this state is
that $R_{\rm t}$ is very close to $R_{\rm co}$, which is also evident
in the tables ($R_{\rm co} / R_*$ = 7.4, 4.6, and 2.9 for $f$ = 0.05,
0.1, and 0.2, respectively).  It appears that, when a stellar wind
torque balances the accretion torque, it may be common (though not
required) for the disc truncation radius to be close to the corotation
radius (unless the spin rate is substantially less than $f = 0.1$).

     \subsection{Accretion Power} \label{sec_power}

In the APSW scenario proposed in Paper~I, the energy that powers the
stellar wind ultimately comes from the energy released by the
accretion process.  In this section, we calculate what fraction of the
accretion power would be required to drive the wind, in the specific
cases for which we have determined the spin-equilibrium.

In order to tabulate the accretion power, the precise details of the
complicated interaction between the star and disk are not important.
The general behavior is that material from the Keplerian disk becomes
attached to the stellar magnetosphere near $R_{\rm t}$ and eventually
falls onto and becomes part of the star.  Energetically, this can be
treated as an inelastic process, wherein only the energy content
before and after the interaction needs to be specified.  Thus, the
rate of potential energy release is simply 1/2 $ \dot M_{\rm a} v_{\rm
  esc}^2 (1 - R_*/R_{\rm t})$.  Note that as accreting material piles
onto the stellar surface, there should be additional energy released
as material either ``sinks" into (convectively) or compresses the
star.  This is another potential energy source, but we neglect this
here.  The rate of (rotational) kinetic energy release is 1/4 $ \dot
M_{\rm a} v_{\rm esc}^2 (R_*/R_{\rm t} - k^2 f^2)$, where the last
term assumes that accreting material eventually achieves the same
specific angular momentum as the star.  The difference in thermal
energy density between material at the disk inner edge and material at
stellar photospheric temperature is negligible compared to the
potential and kinetic energy release.  Thus, by summing the potential
and kinetic energies, the rate of energy release in the vicinity of
the star is approximately
\begin{eqnarray}
\label{eqn_edota}
\dot E_{\rm a} = {1 \over 2} \dot M_{\rm a} v_{\rm esc}^2
       \left({1 - {1 \over 2} {R_* \over R_{\rm t}} - {1 \over 2} k^2 f^2}\right).
%
\end{eqnarray}
Since there is an accretion torque on the star, some of this energy is
added to the rotational energy of the star at a rate $\Omega_*
\tau_{\rm a}$.  The remaining energy ($\dot E_{\rm a} - \Omega_*
\tau_{\rm a}$) is available to power other accretion-related
activity\footnote{We follow a similar derivation of the accretion
  power to that of Paper~I, where equation (4) of that work
  corresponds to the remaining energy, $\dot E_{\rm a} - \Omega_*
  \tau_{\rm a}$, and neglects terms proportional to $f^2$.}.  In
particular, this remaining energy is responsible for powering the
observed excess continuum emission (such as the UV excess) and line
emission \citep[e.g.,][]{konigl91, calvetgullbring98, muzerolle3ea01},
in addition to driving an enhanced stellar wind (Paper~I).

Paper~I proposed that a fraction $\epsilon$ of this energy
specifically powers the thermal energy in the stellar wind.  Our
simulated winds are thermally driven, but the wind driving mechanism
at work in real systems is still uncertain (see Paper~II).  Thus, we
wish to calculate the power required to drive the wind, in a generic
form.  For this, we simply calculate the total energy in the wind far
from the star plus the potential energy required to lift the wind off
the stellar surface.  In this way, the power in a steady-state, 2.5D,
MHD wind can be obtained by \citep[see, e.g.,][]{ustyugovaea99,
  keppensgoedbloed00}
\begin{eqnarray}
\label{eqn_edotw}
\dot E_{\rm w} = 4 \pi R^2 \int_0^1 \rho v_{\rm R} E^\prime d(\cos \theta)
                 + {1 \over 2} \dot M_{\rm w} v_{\rm esc}^2,
\end{eqnarray}
where $\theta$ is the usual spherical coordinate and
\begin{eqnarray}
\label{eqn_e}
E^\prime \equiv {v_{\rm p}^2 + v_\phi^2 \over 2} + {B_\phi^2 \over 4
  \pi \rho} - {v_\phi B_\phi B_{\rm p} \over 4 \pi \rho v_{\rm p}}.
\end{eqnarray}
In equation (\ref{eqn_e}), we have neglected the thermal and
gravitational potential energy, so the integral in equation
(\ref{eqn_edotw}) should be evaluated at large $R$, where $E^\prime$
has reached an asymptotic value and these energies are negligible.
Thus, $\dot E_{\rm w}$ represents the total power required to lift
material off of the star, to accelerate it to the wind velocity, and
to provide the magnetic energy content carried with the wind.

For each of our simulated wind solutions, we evaluate the integral in
equation (\ref{eqn_edotw}) at a radius of $R = 50 R_*$, where
$E^\prime$ is within a few percent of its asymptotic value.  The spin
of the star does work on the wind at a rate $\Omega_* \tau_{\rm w}$.
This represents the power injected in the wind by magnetocentrifugal
processes.  We find that the ratio $\Omega_* \tau_{\rm w} / \dot
E_{\rm w}$ is 30\% in the fiducial case.  In most other cases, the
value of this ratio falls the range 10--60\%.  This indicates that, as
discussed by Paper~I and II \citep[and see][]{washimishibata93}, these
winds are in a regime where the magnetocentrifugal effects are of
nearly equal importance with the other source of wind driving.

It is this other source of wind driving that we propose is powered by
some fraction of the available accretion energy.  We define this
fraction as\footnote{This fraction is a more general definition than
$\epsilon$ in Paper~I, which assumes thermal wind driving. By
contrast, $\epsilon_\infty$ is the fraction of the accretion power
required to explain the energy in the wind at large distances from the
star, regardless of the driving mechanism.}
\begin{eqnarray}
\label{eqn_epsilon}
\epsilon_\infty \equiv {{\dot E_{\rm w} - \Omega_* \tau_{\rm w}} \over 
            {\dot E_{\rm a} - \Omega_* \tau_{\rm a}}}.
\end{eqnarray}
This represents the minimum fraction of the accretion power required
to drive the stellar wind, since whatever mechanism drives the wind
will not itself likely be 100\% efficient \citep{decampli81}.

In the last column of tables \ref{tab_equilib0.1} and
\ref{tab_equilib0.01}, we list the value of $\epsilon_\infty$ for each
case in spin-equilibrium.  In one case ($f = 0.05$), $\epsilon_\infty$
is greater than 100\%, indicating that there is not enough accretion
power in the spin equilibrium state to power the wind.  This case is
therefore not an acceptable solution for a system in spin-equilibrium
by an APSW.  All of the other cases have $\epsilon_\infty < 1$, and so
they are energetically viable solutions.

There is a relationship between $\epsilon_\infty$, the observed excess
emission, and the inferred mass accretion rate.  In particular, the
accretion rates are typically determined by measuring excess emission
and assuming that all of the accretion power is radiated
\citep[e.g.,][]{calvetgullbring98}.  In the APSW scenario, some of the
accretion power drives the stellar wind, so only a fraction $1 -
\epsilon_\infty$ of the accretion power can be radiated.  This means
that the true accretion rate ($\dot M_{\rm a}$) will be a factor of
$(1 - \epsilon_\infty)^{-1}$ larger than the observationally
determined value.  In this context, and since $\epsilon_\infty$ is the
miminum required fraction to drive the wind, a value of
$\epsilon_\infty \ga 0.5$ (as for several cases listed in tables
\ref{tab_equilib0.1} and \ref{tab_equilib0.01}) appears quite large.
However, the observational determination of $\dot M_{\rm a}$ is
uncertain by a factor of several, as exemplified by the large range of
measurements compiled by \citet{johnskrullgafford02}.  Thus, while it
is clear that $\epsilon_\infty < 1$ is a hard upper limit, it is not
yet clear how close to unity $\epsilon_\infty$ can be.

As expected, the cases with lower values of $\dot M_{\rm w} / \dot
M_{\rm a}$ (i.e., cases with larger field strength or smaller $\dot
M_{\rm w}$) require a smaller fraction of the accretion power to drive
the wind.  The cases in the table suggest approximately that
$\epsilon_\infty \approx 1.6 \dot M_{\rm w} / \dot M_{\rm a}$.  Thus,
it appears that $\dot M_{\rm w} / \dot M_{\rm a} \la 0.6$ represents a
hard upper limit for APSWs.

It is important to note that the manner in which the accretion power
transfers to the stellar wind is still unspecified in the APSW model.
This will depend upon what is the wind driving mechanism, which is
currently unknown.  In reality, the physics of the energy coupling
will likely determine the value of $\epsilon_\infty$, which
effectively sets the value for $\dot M_{\rm w} / \dot M_{\rm a}$.
Then, given enough time, the stellar spin rate will evolve toward the
equilibrium value set primarily by $\dot M_{\rm w} / \dot M_{\rm a}$
and $r_{\rm A}/R_*$.  Thus the spin-equilibrium state of the star is
ultimately determined by the power coupling and magnetic properties,
and more work is needed to take this further.

     \subsection{Semi-Analytic Formulation for Spin Equilibrium} 

\label{sec_semianalytic}

In order to develop a more general, predictive theory, in this section
we make use of the semi-analytic formulation of equation
(\ref{eqn_rasim}).  As justified by the previous sections, we assume
that the equilibrium spin rate of the star is simply determined by a
balance between the spin-up torque from accretion and the spin-down
torque from the stellar wind.  Using equations (\ref{eqn_f}) --
(\ref{eqn_rasim}), (\ref{eqn_ta}), and (\ref{eqn_teq}), we can write
the stellar wind equilibrium spin rate, expressed as a fraction of
breakup spin, as
\begin{eqnarray}
\label{eqn_fsw}
%
f_{\rm sw} = K^{-2}
               \left({R_{\rm t} \over R_*}\right)^{1/2}
               \left({\dot M_{\rm a} \over \dot M_{\rm w}}\right)^{1-2m}
	       \Psi^{-2m},
\end{eqnarray}
where
\begin{eqnarray}
\label{eqn_psi}
\Psi \equiv  {{B_*^2 R_*^2} \over {\dot M_{\rm a} v_{\rm esc}}}
\end{eqnarray}
is a dimensionless magnetization parameter\footnote{The magnetization
  parameter $\Psi$ is related to $\psi$ used by MP05b (see their
  eq.\ 16) by a constant factor, $\psi = 2^{3/2} \Psi$.}.  Here, we
have neglected the term proportional to $k^2 f$ in equation
(\ref{eqn_ta}), since it is generally much smaller than the other
term.  Again, note that a weak $f$-dependence of $r_{\rm A}$ is not
included in equation (\ref{eqn_rasim}), so the dependence of $f_{\rm
  sw}$ on some of the parameters is not precisely captured in equation
(\ref{eqn_fsw}) (see Paper~II).

Equation (\ref{eqn_fsw}) includes a dependence on the truncation
radius of the disk, $R_{\rm t}$.  This location itself has a
dependence on the other parameters, and the determination of $R_{\rm
t}$ depends on the magnetic connection state of the system (\S
\ref{sec_state}; MP05b).  In general, $R_{\rm t}$ depends on $\Psi$,
but if $R_{\rm t}$ is close to the corotation radius, $R_{\rm co}$,
then $R_{\rm t}$ also depends on the stellar spin rate.  We will
consider two cases that are expected to bracket reality.

The first case is one in which the system is in state 1 as defined by
MP05b.  Here, $R_{\rm t}$ does not depend on the stellar spin rate,
and it is simply proportional to the original calculations by
\citet{lamb3ea73} and \citet{davidsonostriker73} used in most Ghosh \&
Lamb type models.  Thus, using equation (\ref{eqn_rt1}) for $R_{\rm t}$,
adopting $\gamma_{\rm c} = 1$, and plugging in to equation
(\ref{eqn_fsw}), one finds
\begin{eqnarray}
\label{eqn_fsw1}
f_{\rm sw1} = {{2^{3/14}} \over {K^{2}}}
               \left({\dot M_{\rm a} \over \dot M_{\rm w}}\right)^{1-2m}
	       \Psi^{1/7-2m}.
\end{eqnarray}
This is the predicted equilibrium spin rate when the truncation radius
is significantly smaller than the corotation radius (i.e., in magnetic
connection state 1).

The second case to consider is where $R_{\rm t} \approx R_{\rm co}$,
which is the requirement of all disk-locking models
\citep[][MP05b]{shuea94, wang95}.  Using equation (\ref{eqn_rco}) and
setting $R_{\rm t} = R_{\rm co}$ in equation (\ref{eqn_fsw}), one
finds
\begin{eqnarray}
\label{eqn_fsw2}
f_{\rm sw2} = K^{-3/2}
               \left({\dot M_{\rm a} \over \dot M_{\rm w}}\right)^{(3-6m)/4}
	       \Psi^{-3m/2}.
\end{eqnarray}
This is the predicted equilibrium spin rate when the disk truncation
occurs very close to $R_{\rm co}$.

Which case is more appropriate?  The first case is expected to occurs
for relatively small values of $\Psi$ and low spin rates $f$ (MP05b),
and the opposite is true for the second case.  For the parameter space
we have considered thus far in this work, we found in section
\ref{sec_state} that most (though not all) of the cases are expected
to have $R_{\rm t} / R_{\rm co}$ near unity.  Thus, while it is not a
formal requirement of APSW, it may often be the case that $R_{\rm t}
\approx R_{\rm co}$ for systems in spin equilibrium, and we will focus
on this second case for the remainder of this work.

\begin{figure}
\epsscale{1.15}
\plotone{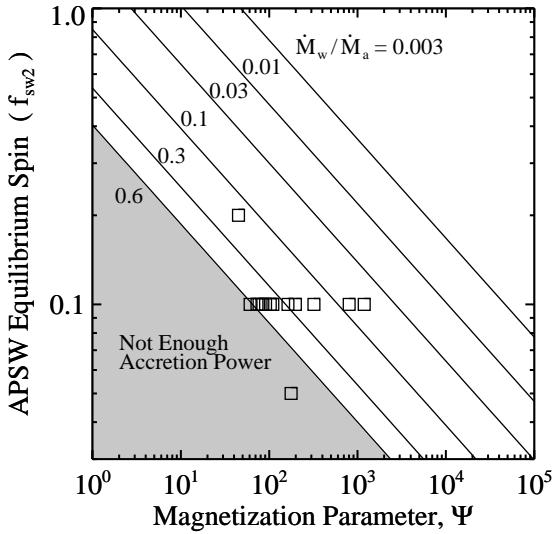}

\caption{The equilibrium spin rate predicted by a balance between the
spin down from a stellar wind and the spin up from accretion, versus
the dimensionless magnetization parameter $\Psi \equiv B_*^2 R_*^2
(\dot M_{\rm a} v_{\rm esc})^{-1}$.  The solid lines show equation
(\ref{eqn_fsw2}) for $K = 2.11$, $m = 0.223$ and several different
possible values of $\dot M_{\rm w} / \dot M_{\rm a}$, as indicated.
The squares show data from table \ref{tab_equilib0.1}, which indicates
the range of parameters considered in our simulations and used to
derive equation (\ref{eqn_fsw2}).  The shaded region corresponds
approximately to where accretion power is not sufficient to drive the
stellar wind.
\label{fig_feqs1}}

\end{figure}

Figure \ref{fig_feqs1} shows the predicted spin rate of this case
(eq.\ \ref{eqn_fsw2}) versus $\Psi$, for many different values of the
ratio $\dot M_{\rm w} / \dot M_{\rm a}$.  The results of section
\ref{sec_power} indicate that the accretion power is only capable of
powering a wind with $\dot M_{\rm w} / \dot M_{\rm a} \la 0.6$, and
this is a hard upper limit.  This ``forbidden'' region of the
$f$-$\Psi$ space is indicated in figure \ref{fig_feqs1}.

The plot also shows the simulation results (squares).  The value of
$\Psi$ for each case is determined mostly by input parameters but also
by $\dot M_{\rm a}$.  The latter was set (in section
\ref{sec_balancing}) by the condition that the equilibrium spin rate
was equal to the value of $f$ used as the simulation input parameter.
The squares indicate the range over which equation (\ref{eqn_fsw2}) is
shown to be valid by the simulations.  Also, the plot shows the
specific case of $K = 2.11$ and $m = 0.223$.

The results shown in figure \ref{fig_feqs1} and equation
(\ref{eqn_fsw2}) provide the basis for predictions of the APSW model
that can be observationally tested and constrained, and that can be
compared to other models.  Different theories predict different power
laws for $f$ vs.\ $\Psi$.  Specifically, equation (\ref{eqn_fsw2})
predicts a power law index of $\approx -0.33$, the conditions
appropriate for equation (\ref{eqn_fsw1}) predict $\approx -0.30$, and
an index of $-3/7$ is predicted by the disk locking models
\citep[e.g.,][]{konigl91, shuea94}.



\section{Summary and Discussion} \label{sec_discussion}

In this work, we have further developed the accretion-powered stellar
wind model proposed in Paper~I, where the stellar wind magnetic lever
arm length, $r_{\rm A}$, was taken as a parameter to determine the
stellar wind torque.  We employed the simulation results of Paper~II
(see \S \ref{sec_simulations}) to obtain stellar wind torques for
several cases representative of T Tauri systems.  We examined the
total torque on the star arising from the stellar wind plus the
magnetic interaction between the star and its accretion disk.  Our
results can be summarized as follows.
\begin{enumerate}

\item{We found that the spin-down torque from a stellar wind can be
  orders of magnitude stronger than the spin-down portion of the
  star-disk interaction torque, for slowly rotating stars with mass
  loss rates substantially larger than the solar wind outflow rate
  (see \S \ref{sec_tdsd}).  This confirms the assumption of Paper~I
  that the condition for net zero torque on the star (spin
  equilibrium) is simply determined by a balance between the stellar
  wind torque and the accretion torque.}

\item{Using the computed stellar wind torques for several cases, we
  looked at the conditions for spin equilibrium (\S
  \ref{sec_balancing}).  We found that a rotation rate of 10\% of
  breakup speed typically requires $\dot M_{\rm w} / \dot M_{\rm a}$
  equal a few tens of percent, confirming the original suggestion by
  \citet{hartmannstauffer89} that stellar winds may be capable of
  removing accreted angular momentum.}

\item{For most cases in spin equilibrium, the disk truncation radius
  was very close to the corotation radius, though this is not a
  general requirement of the APSW model (\S \ref{sec_state}).}

\item{Accretion power is generally sufficient to power a stellar wind
  that is capable of solving the angular momentum problem (\S
  \ref{sec_power}), as suggested in Paper~I.  The energy requirements
  for most of the cases considered here is relatively large, and more
  work is needed to further constrain the energy coupling.}

\item{Under the assumption that the stellar wind is accretion powered,
  the cases we examined suggested a hard upper limit of $\dot M_{\rm
    w} / \dot M_{\rm a} \la 0.6$.}

\item{Finally, in section \ref{sec_semianalytic} we used the results
  from Paper~II to derive a semi-analytic formulation for the
  equilibrium spin rate predicted by the APSW model.  We found the
  that spin rate, expressed as a fraction of breakup speed, generally
  depends only on the two dimensionless parameters $\dot M_{\rm w} /
  \dot M_{\rm a}$ and $\Psi \equiv B_*^2 R_*^2 (\dot M_{\rm a} v_{\rm
    esc})^{-1}$.}

\end{enumerate}

The APSW model incorporates several previous ideas.  As in all other
models that emphasize the role of stellar magnetic fields, the
interaction of the magnetized star with the disk leads to the
truncation of the disk and accretion of material along field lines
onto the star.  The APSW model adopts the finding of the Ghosh \&
Lamb-type models \citep[e.g.,][]{ghoshlamb78, konigl91,
  armitageclarke96}, which is also supported by numerical simulations
\citep[][]{romanovaea02, long3ea05}, that the angular momentum of
accreting material is transferred to the star.  However, in contrast
to the Ghosh \& Lamb-type models, we found that for slow rotators, any
spin-down torque arising from the star-disk interaction is negligible
(item 1 above). Instead, the stellar spin-up torque from accretion is
counteracted by an accretion-driven stellar wind, which carries a
comparable amount of angular momentum out of the system.

Compared to all existing models, APSW is distinct in that it
conceptually links the driving of the stellar wind to the energy
released by the accretion process (via $\epsilon$).  In other ways,
the general picture of APSW is similar to other angular momentum
models that utilize winds.  In particular, the X-wind \citep{shuea94,
  ostrikershu95}, the Reconnection X-wind \citep{ferreira3ea00}, and
other works considering stellar winds \citep{hartmannstauffer89,
  paatzcamenzind96} all find that, in order to carry away significant
angular momentum, the mass outflow rate needs to be of the order of
10\% of the accretion rate (item 2 above).  Except for the X-wind, in
all of the above scenarios, the outflow is magnetically connected to
the star, and thus extracts angular momentum directly from the star.
By contrast, the X-wind outflow is magnetically connected to the disk.
Furthermore, the X-wind is unique in that it assumes that the
accretion of material does not deposit angular momentum onto the star.

As mentioned in item 3 above, for most of the specific cases of our
simulated winds, we found that in the spin equilibrium state, the
truncation radius was very close to the corotation radius.  This is
similar to the prediction of the disk locking models, which includes
both the Ghosh \& Lamb-type models and the X-wind.  However, in
contrast with the disk locking models, it is not a requirement of APSW
that $R_{\rm t}$ be close to $R_{\rm co}$.  Measurements of the
location of the inner edge of the gas disk \citep{najita3ea03, carr07}
suggest that $R_{\rm t}$/$R_{\rm co}$ is typically $\sim 70$\%.  We
leave a more detailed comparison between models for future work.


There is observational evidence that the outflow rates from accreting
young stellar systems are of the order of 10\% of the accretion rates
\citep[e.g.,][]{hartigan3ea95, calvet97} and are therefore accretion
powered \citep{cabritea90}.  However, it appears that a large fraction
of this flow (which is usually probed by forbidden emission coming
from large spatial scales) originates in the disk, rather than the
star \citep{ferreira3ea06}.  It is not yet clear how much of the total
observed flow may originate in a stellar wind.  There is some evidence
specifically for stellar winds from CTTSs \citep{beristain3ea01,
edwardsea03, dupreeea05, edwardsea06, kwan3ea07}, as distinct from
disk winds, and that these are accretion powered
\citep[e.g.,][]{edwardsea03, edwardsea06}, but the mass outflow rates
are not yet well constrained \citep{dupreeea05}.  Additional work
constraining the value of $\dot M_{\rm w} / \dot M_{\rm a}$, the
stellar wind driving mechanism, and the stellar magnetic field
strength and geometry will help to provide stringent and quantitative
tests for the APSW model.

The predictions of the spin equilibrium state can also be checked
observationally \citep{johnskrullgafford02}.  This will likely require
large samples of stars, due to large uncertainties in measured
parameters, and since intrinsic variability in real systems
\citep[e.g.,]{hartmann97} may only allow a spin equilibrium state to
be achieved in a time-averaged sense.  The Ghosh \& Lamb, X-wind, and
APSW models all predict an equilibrium spin rate that depends on
$\Psi$, but of these three, only the APSW model contains an additional
dependence on the stellar wind mass outflow rate.  For the power law
fits to the simulations of Paper~II, the APSW spin equilibrium
predicts a slightly different power-law of spin vs.\ $\Psi$ than the
other models (\S \ref{sec_semianalytic})---though the exact dependence
of the stellar wind torque has not been determined for all parameters.

In this series of papers, we have focused on the global problem of
calculating the magnitude of stellar wind torques and comparing them
with other torques acting on accreting stars.  In order to refine the
APSW model further and make the predictions more precise, more work is
required.  In particular, it is not yet clear how the presence of an
accretion disk will influence the stellar wind torque, and conversely,
how a stellar wind may influence the accretion process.  Also,
although it is clear that there is enough accretion energy to power
the stellar wind, it is still not known what actually drives the
stellar wind and how the accretion power may transfer to it.  We
suspect that a strong flux of hydromagnetic waves can be excited near
the base of the accretion shock and can tap the energy released there,
which may provide an efficient driver for the APSW.  We defer a
rigourous investigation of the APSW driving mechanism to future work.

\acknowledgements

We wish to thank many people for discussions regarding this work,
including: Gibor Basri, Sylvie Cabrit, Andrea Dupree, Suzan Edwards,
Will Fischer, Shu-ichiro Inutsuka, Chris Johns-Krull, Marina Romanova,
Frank Shu, Keivan Stassun, Jeff Valenti, and Sydney Wolff.  We also
thank the referee, Jonathan Ferreira, for his useful suggestions for
improving the paper and KITP for hosting us while finishing the
manuscript.  This research was supported in part by the National
Science Foundation under Grant No. PHY05-51164.  SM is supported by
the University of Virginia through a Levinson/VITA Fellowship
partially funded by The Frank Levinson Family Foundation through the
Peninsula Community Foundation.  REP is supported by a grant from
NSERC.

\appendix
\section{Determination of the Disk Truncation Radius}

We follow MP05b to calculate the location of the disk truncation
radius, $R_{\rm t}$, and the reader will find details in that paper.
For convenience, we list the relevant equations here.  As in section
\ref{sec_tdsd}, we adopt $\gamma_{\rm c} = 1$.

To determine $R_{\rm t}$, we first find the magnetic connectivity
state using the criterion
\begin{eqnarray}
\label{eqn_trans}
f < (1 - \beta) (2^{3/2} \Psi)^{-3/7},
\end{eqnarray}
where $\Psi$ is defined by equation (\ref{eqn_psi}).  If condition
(\ref{eqn_trans}) is satisfied, the system is in ``state 1."
Otherwise, it is in ``state 2."  In state 1, we determine the
truncation radius using
\begin{eqnarray}
\label{eqn_rt1}
R_{\rm t} = (2^{3/2} \Psi)^{2/7} R_*.
\end{eqnarray}
In state 2, we determine the truncation radius by solving
\begin{eqnarray}
\label{eqn_rt2}
\left({R_{\rm t} \over R_{\rm co}}\right)^{-7/2}
  \left[{1 - \left({R_{\rm t} \over R_{\rm co}}\right)^{3/2}}\right] =
  {\beta \over 2^{3/2} \Psi} f^{-{7/3}}.
\end{eqnarray}



\begin{thebibliography}{56}
\expandafter\ifx\csname natexlab\endcsname\relax\def\natexlab#1{#1}\fi

\bibitem[{{Armitage} \& {Clarke}(1996)}]{armitageclarke96}
{Armitage}, P.~J. \& {Clarke}, C.~J. 1996, \mnras, 280, 458

\bibitem[{{Beristain} {et~al.}(2001){Beristain}, {Edwards}, \&
  {Kwan}}]{beristain3ea01}
{Beristain}, G., {Edwards}, S., \& {Kwan}, J. 2001, \apj, 551, 1037

\bibitem[{{Bertout} {et~al.}(1988){Bertout}, {Basri}, \&
  {Bouvier}}]{bertout3ea88}
{Bertout}, C., {Basri}, G., \& {Bouvier}, J. 1988, \apj, 330, 350

\bibitem[{{Bouvier} {et~al.}(1997){Bouvier}, {Forestini}, \&
  {Allain}}]{bouvier3ea97}
{Bouvier}, J., {Forestini}, M., \& {Allain}, S. 1997, \aap, 326, 1023

\bibitem[{{Cabrit} {et~al.}(1990){Cabrit}, {Edwards}, {Strom}, \&
  {Strom}}]{cabritea90}
{Cabrit}, S., {Edwards}, S., {Strom}, S.~E., \& {Strom}, K.~M. 1990, \apj, 354,
  687

\bibitem[{{Calvet}(1997)}]{calvet97}
{Calvet}, N. 1997, in IAU Symposium, Vol. 182, Herbig-Haro Flows and the Birth
  of Stars, ed. B.~{Reipurth} \& C.~{Bertout}, 417--432

\bibitem[{{Calvet} \& {Gullbring}(1998)}]{calvetgullbring98}
{Calvet}, N. \& {Gullbring}, E. 1998, \apj, 509, 802

\bibitem[{{Camenzind}(1990)}]{camenzind90}
{Camenzind}, M. 1990, in Reviews in Modern Astronomy, ed. G.~{Klare}, 234--265

\bibitem[{{Cameron} \& {Campbell}(1993)}]{cameroncampbell93}
{Cameron}, A.~C. \& {Campbell}, C.~G. 1993, \aap, 274, 309

\bibitem[{{Carr}(2007)}]{carr07}
{Carr}, J.~S. 2007, to appear in proceedings of IAU Symposium No.\ 243,
  Star-Disk Interaction in Young Stars

\bibitem[{{Davidson} \& {Ostriker}(1973)}]{davidsonostriker73}
{Davidson}, K. \& {Ostriker}, J.~P. 1973, \apj, 179, 585

\bibitem[{{Decampli}(1981)}]{decampli81}
{Decampli}, W.~M. 1981, \apj, 244, 124

\bibitem[{{Dupree} {et~al.}(2005){Dupree}, {Brickhouse}, {Smith}, \&
  {Strader}}]{dupreeea05}
{Dupree}, A.~K., {Brickhouse}, N.~S., {Smith}, G.~H., \& {Strader}, J. 2005,
  \apjl, 625, L131

\bibitem[{{Edwards} {et~al.}(2006){Edwards}, {Fischer}, {Hillenbrand}, \&
  {Kwan}}]{edwardsea06}
{Edwards}, S., {Fischer}, W., {Hillenbrand}, L., \& {Kwan}, J. 2006, \apj, 646,
  319

\bibitem[{{Edwards} {et~al.}(2003){Edwards}, {Fischer}, {Kwan}, {Hillenbrand},
  \& {Dupree}}]{edwardsea03}
{Edwards}, S., {Fischer}, W., {Kwan}, J., {Hillenbrand}, L., \& {Dupree}, A.~K.
  2003, \apjl, 599, L41

\bibitem[{{Ferreira}(1997)}]{ferreira97}
{Ferreira}, J. 1997, \aap, 319, 340

\bibitem[{{Ferreira} {et~al.}(2006){Ferreira}, {Dougados}, \&
  {Cabrit}}]{ferreira3ea06}
{Ferreira}, J., {Dougados}, C., \& {Cabrit}, S. 2006, \aap, 453, 785

\bibitem[{{Ferreira} {et~al.}(2000){Ferreira}, {Pelletier}, \&
  {Appl}}]{ferreira3ea00}
{Ferreira}, J., {Pelletier}, G., \& {Appl}, S. 2000, \mnras, 312, 387

\bibitem[{{Ghosh} \& {Lamb}(1978)}]{ghoshlamb78}
{Ghosh}, P. \& {Lamb}, F.~K. 1978, \apjl, 223, L83

\bibitem[{{Hartigan} {et~al.}(1995){Hartigan}, {Edwards}, \&
  {Ghandour}}]{hartigan3ea95}
{Hartigan}, P., {Edwards}, S., \& {Ghandour}, L. 1995, \apj, 452, 736

\bibitem[{{Hartmann}(1997)}]{hartmann97}
{Hartmann}, L. 1997, in IAU Symposium, Vol. 182, Herbig-Haro Flows and the
  Birth of Stars, ed. B.~{Reipurth} \& C.~{Bertout}, 391--405

\bibitem[{{Hartmann} \& {Stauffer}(1989)}]{hartmannstauffer89}
{Hartmann}, L. \& {Stauffer}, J.~R. 1989, \aj, 97, 873

\bibitem[{{Herbst} {et~al.}(2007){Herbst}, {Eisl{\"o}ffel}, {Mundt}, \&
  {Scholz}}]{herbstea07}
{Herbst}, W., {Eisl{\"o}ffel}, J., {Mundt}, R., \& {Scholz}, A. 2007, in
  Protostars and Planets V, ed. B.~{Reipurth}, D.~{Jewitt}, \& K.~{Keil},
  297--311

\bibitem[{{Jayawardhana} {et~al.}(2006){Jayawardhana}, {Coffey}, {Scholz},
  {Brandeker}, \& {van Kerkwijk}}]{jayawardhanaea06}
{Jayawardhana}, R., {Coffey}, J., {Scholz}, A., {Brandeker}, A., \& {van
  Kerkwijk}, M.~H. 2006, \apj, 648, 1206

\bibitem[{{Johns-Krull} \& {Gafford}(2002)}]{johnskrullgafford02}
{Johns-Krull}, C.~M. \& {Gafford}, A.~D. 2002, \apj, 573, 685

\bibitem[{{Joy}(1945)}]{joy45}
{Joy}, A.~H. 1945, \apj, 102, 168

\bibitem[{{Keppens} \& {Goedbloed}(2000)}]{keppensgoedbloed00}
{Keppens}, R. \& {Goedbloed}, J.~P. 2000, \apj, 530, 1036

\bibitem[{{K\"onigl}(1991)}]{konigl91}
{K\"onigl}, A. 1991, \apjl, 370, L39

\bibitem[{{Kwan} {et~al.}(2007){Kwan}, {Edwards}, \& {Fischer}}]{kwan3ea07}
{Kwan}, J., {Edwards}, S., \& {Fischer}, W. 2007, \apj, 657, 897

\bibitem[{{Lamb} {et~al.}(1973){Lamb}, {Pethick}, \& {Pines}}]{lamb3ea73}
{Lamb}, F.~K., {Pethick}, C.~J., \& {Pines}, D. 1973, \apj, 184, 271

\bibitem[{{Long} {et~al.}(2005){Long}, {Romanova}, \& {Lovelace}}]{long3ea05}
{Long}, M., {Romanova}, M.~M., \& {Lovelace}, R.~V.~E. 2005, \apj, 634, 1214

\bibitem[{{Lynden-Bell} \& {Pringle}(1974)}]{lyndenbellpringle74}
{Lynden-Bell}, D. \& {Pringle}, J.~E. 1974, \mnras, 168, 603

\bibitem[{{Lyo} \& {Lawson}(2005)}]{lyolawson05}
{Lyo}, A.-R. \& {Lawson}, W.~A. 2005, Journal of Korean Astronomical Society,
  38, 241

\bibitem[{{Matt} \& {Pudritz}(2005{\natexlab{a}})}]{mattpudritz05l}
{Matt}, S. \& {Pudritz}, R.~E. 2005{\natexlab{a}}, \apjl, 632, L135 (Paper~I)

\bibitem[{{Matt} \& {Pudritz}(2005{\natexlab{b}})}]{mattpudritz05}
---. 2005{\natexlab{b}}, \mnras, 356, 167 (MP05b)

\bibitem[{{Matt} \& {Pudritz}(2007)}]{mattpudritz07coolstars}
---. 2007, to appear in proceedings of the 14th Cambridge Workshop on Cool
  Stars, Stellar Systems, and the Sun, astro-ph/0701648

\bibitem[{{Matt} \& {Pudritz}(2008)}]{mattpudritz08II}
---. 2008, \apj, in press (arXiv:0801.0436) (Paper~II)

\bibitem[{{Muzerolle} {et~al.}(2001){Muzerolle}, {Calvet}, \&
  {Hartmann}}]{muzerolle3ea01}
{Muzerolle}, J., {Calvet}, N., \& {Hartmann}, L. 2001, \apj, 550, 944

\bibitem[{{Najita} {et~al.}(2003){Najita}, {Carr}, \& {Mathieu}}]{najita3ea03}
{Najita}, J., {Carr}, J.~S., \& {Mathieu}, R.~D. 2003, \apj, 589, 931

\bibitem[{{Ostriker} \& {Shu}(1995)}]{ostrikershu95}
{Ostriker}, E.~C. \& {Shu}, F.~H. 1995, \apj, 447, 813

\bibitem[{{Paatz} \& {Camenzind}(1996)}]{paatzcamenzind96}
{Paatz}, G. \& {Camenzind}, M. 1996, \aap, 308, 77

\bibitem[{{Pelletier} \& {Pudritz}(1992)}]{pelletierpudritz92}
{Pelletier}, G. \& {Pudritz}, R.~E. 1992, \apj, 394, 117

\bibitem[{{Rebull} {et~al.}(2004){Rebull}, {Wolff}, \& {Strom}}]{rebull3ea04}
{Rebull}, L.~M., {Wolff}, S.~C., \& {Strom}, S.~E. 2004, \aj, 127, 1029

\bibitem[{{Rebull} {et~al.}(2002){Rebull}, {Wolff}, {Strom}, \&
  {Makidon}}]{rebullea02}
{Rebull}, L.~M., {Wolff}, S.~C., {Strom}, S.~E., \& {Makidon}, R.~B. 2002, \aj,
  124, 546

\bibitem[{{Romanova} {et~al.}(2002){Romanova}, {Ustyugova}, {Koldoba}, \&
  {Lovelace}}]{romanovaea02}
{Romanova}, M.~M., {Ustyugova}, G.~V., {Koldoba}, A.~V., \& {Lovelace},
  R.~V.~E. 2002, \apj, 578, 420

\bibitem[{{Safier}(1998)}]{safier98}
{Safier}, P.~N. 1998, \apj, 494, 336

\bibitem[{{Shakura} \& {Sunyaev}(1973)}]{shakurasunyaev73}
{Shakura}, N.~I. \& {Sunyaev}, R.~A. 1973, \aap, 24, 337

\bibitem[{{Shu} {et~al.}(1994){Shu}, {Najita}, {Ostriker}, {Wilkin}, {Ruden},
  \& {Lizano}}]{shuea94}
{Shu}, F., {Najita}, J., {Ostriker}, E., {Wilkin}, F., {Ruden}, S., \&
  {Lizano}, S. 1994, \apj, 429, 781

\bibitem[{{Tout} \& {Pringle}(1992)}]{toutpringle92}
{Tout}, C.~A. \& {Pringle}, J.~E. 1992, \mnras, 256, 269

\bibitem[{{Ustyugova} {et~al.}(1999){Ustyugova}, {Koldoba}, {Romanova},
  {Chechetkin}, \& {Lovelace}}]{ustyugovaea99}
{Ustyugova}, G.~V., {Koldoba}, A.~V., {Romanova}, M.~M., {Chechetkin}, V.~M.,
  \& {Lovelace}, R.~V.~E. 1999, \apj, 516, 221

\bibitem[{{Uzdensky}(2004)}]{uzdensky04}
{Uzdensky}, D.~A. 2004, \apss, 292, 573

\bibitem[{{Uzdensky} {et~al.}(2002){Uzdensky}, {K{\" o}nigl}, \&
  {Litwin}}]{uzdensky3ea02}
{Uzdensky}, D.~A., {K{\" o}nigl}, A., \& {Litwin}, C. 2002, \apj, 565, 1191

\bibitem[{{Vogel} \& {Kuhi}(1981)}]{vogelkuhi81}
{Vogel}, S.~N. \& {Kuhi}, L.~V. 1981, \apj, 245, 960

\bibitem[{{Wang}(1995)}]{wang95}
{Wang}, Y.-M. 1995, \apjl, 449, L153

\bibitem[{{Washimi} \& {Shibata}(1993)}]{washimishibata93}
{Washimi}, H. \& {Shibata}, S. 1993, \mnras, 262, 936

\bibitem[{{Yi}(1995)}]{yi95}
{Yi}, I. 1995, \apj, 442, 768

\end{thebibliography}




\label{lastpage}
\end{document}